\documentclass[
reprint,
superscriptaddress,
amsmath,amssymb,
aps,
prx,
longbibliography
]{revtex4-1}

\usepackage{graphicx}% include figure files
\usepackage{dcolumn}% align table columns on decimal point
\usepackage{bm}% bold math
\usepackage[colorlinks=true,urlcolor=blue,linkcolor=blue,citecolor=blue]{hyperref}% add hypertext capabilities
\usepackage{ulem}

\newcommand{\customref}[2]{\hyperref[#1]{\ref*{#1}#2}}

\usepackage{xcolor}

\begin{document}

\title{Kinetically constrained freezing transition in a dipole-conserving system}

\author{Alan Morningstar}
\affiliation{Department of Physics, Princeton University, Princeton, New Jersey 08544, USA}
 
\author{Vedika Khemani}
\affiliation{Department of Physics, Stanford University, Stanford, CA 94305, USA}

\author{David A. Huse}
\affiliation{Department of Physics, Princeton University, Princeton, New Jersey 08544, USA}
\affiliation{Institute for Advanced Study, Princeton, NJ 08540, USA}

\date{\today}

\begin{abstract}

We study a stochastic lattice gas of particles in one dimension with strictly finite-range interactions that respect the fracton-like conservation laws of total charge and dipole moment.  As the charge density is varied, the connectivity of the system's charge configurations under the dynamics changes qualitatively. We find two distinct phases: Near half filling the system thermalizes subdiffusively, with almost all configurations belonging to a single dynamically connected sector. As the charge density is tuned away from half filling there is a phase transition to a frozen phase where locally active finite bubbles cannot exchange particles and the system fails to thermalize.  The two phases exemplify what has recently been referred to as weak and strong \textit{Hilbert space fragmentation}, respectively.  We study the static and dynamic scaling properties of this weak-to-strong fragmentation phase transition in a kinetically constrained classical Markov circuit model, obtaining some conjectured exact critical exponents.

\end{abstract}

\maketitle

\section{\label{sec:intro} Introduction}

The exploration of far-from-equilibrium quantum many-body systems continues to be a rich source of interesting new phenomena. While many systems eventually approach thermal equilibrium~\cite{Deutsch1991,Srednicki1994,Rigol-Olshanii2009,Dalessio-Rigol2016}, understanding how and when thermalization breaks down in a many-body system is of fundamental interest. At least one robust mechanism for avoiding thermalization has been established: many-body localization (MBL)~\cite{Anderson58, Basko-Altshuler2006, Pal-Huse2010, Oganesyan-Huse2007, Prelovsek, Imbrie2016, Nandkishore-Huse2015, mblrmp}, and along with it has come an active effort to understand the associated dynamical phase transition, where the MBL mechanism breaks down and the system thermalizes~\cite{Pal-Huse2010, Alet-Laflorencie2018, Kjall14,SerbynCriterion, KhemaniCP, ClarkBimodal, VHA, PVP, ZhangRG, DVP, Thiery-DeRoeck2018, GVS, Dumitrescu-Vasseur2019, Morningstar-Huse2019,Gopalakrishnan-Huse2019}. While MBL systems rely on the quenched breaking of translational invariance to induce localization, the search for mechanisms to circumvent thermalization in translationally-invariant systems is also an active area of research~\cite{Grover-Fisher2014,Schiulaz-Muller2015,vanHorssen-Garrahan2015,Yao-Moore2016,Hickey-Garrahan2016,Smith-Moessner2017,Brenes-Scardicchio2018,vanNieuwenburg-Rafael2019,Schulz-Pollmann2019}. Among recent developments in this direction is the exploration of constrained quantum systems such as Rydberg-blockaded chains~\cite{Bernien-Lukin2017}, which has lead to a surge of research on nonthermal ``quantum scar" states~\cite{ShiraishiMori, AKLT1, Turner-Papic2018a,Turner-Papic2018b,Ho-Lukin2019,Khemani-Chandran2019,Lin-Motrunich2019,Pancotti-Banuls2019,SchecterXY,Iadecola,LesikTower, SanjayTower,Moudgalya-Regnault2019,Moudgalya-Bernevig2020,Mark-Motrunich2020a}, and models with fracton-like excitations~\cite{Chamon2005,Haah2011,Vijay-Fu2015,Nandkishore-Hermele2019,Pretko-You2020}, which have been shown to exhibit a form of localization under certain ideal conditions~\cite{Khemani-Nandkishore2019,Sala-Pollmann2019,Rakovszky-Pollmann2019,Moudgalya-Bernevig2019}. In this work we explore a dynamical phase transition between thermalizing and frozen phases that occurs in one such system with fracton-like constraints on the dynamics.

Refs.~\cite{Khemani-Nandkishore2019,Sala-Pollmann2019,Rakovszky-Pollmann2019,Moudgalya-Bernevig2019} showed that the combination of strictly finite-range interactions and the fracton-like constraints of charge and dipole conservation results in a \textit{fragmentation} of Hilbert space into exponentially many (in volume) dynamically disconnected sectors, herein referred to as \textit{Krylov sectors}. This means that a graph, where nodes represent charge configurations and edges represent local dipole-conserving transitions, consists of exponentially many disconnected components $\textit{within}$ each of the polynomially many symmetry sectors of configurations with a common total charge and total dipole moment. Hilbert space fragmentation comes in two distinct types: strong and weak~\cite{Khemani-Nandkishore2019,Sala-Pollmann2019}. In systems that are \textit{strongly} fragmented, the system fails to thermalize because for any initial charge configuration it is constrained to explore only a vanishingly small fraction of the states with the same charge and dipole moment.  On the other hand, in the case of \textit{weak} fragmentation the configuration space still shatters into exponentially many disconnected Krylov sectors, but the fraction of states that belong to the largest sector approaches one in the limit of large systems, and therefore typical initial states are in this largest Krylov sector and thus can thermalize. 

In this paper we study a charge and dipole-conserving model that is weakly fragmented at charge densities around half filling, for which typical initial states do thermalize.  However, as the total charge in the system is varied, a critical point is encountered beyond which thermalization cannot occur due to strong fragmentation of the set of all charge configurations with that total charge. Since the fragmentation in this case is dictated by fundamentally classical constraints---because we are working in a basis where fractonic degrees of freedom are assumed to exist without having to emerge from an underlying quantum model---we study a kinetically constrained~\cite{Ritort-Sollich2003,Garrahan-Toninelli2010} classical Markov circuit model. We study the static and dynamic properties of the phase transition between the weakly and strongly fragmented phases by constructing and supporting a simplified theoretical model that we believe captures many of the key features of the critical point, as well as by simulating the full dynamics.  Our simplified model gives conjectured exact critical exponents of $\nu=2$ for the correlation length, and $\beta=1$ for the density of frozen sites which serves as an ``order parameter'' for the frozen phase.  
The dynamics in the thermalizing, weakly-fragmented phase away from the phase transition is demonstrated to be subdiffusive, with the transport time growing as the fourth power of the length. At the phase transition, the dynamics is slower, as expected, with a dynamical critical exponent $z$ that appears to be near $7$, but this can only be explored numerically over a modest range of length scales, so the true asymptotic $z$ might be larger than this.

\section{\label{sec:model} Model}

\begin{figure}
\includegraphics[width=1.0\linewidth]{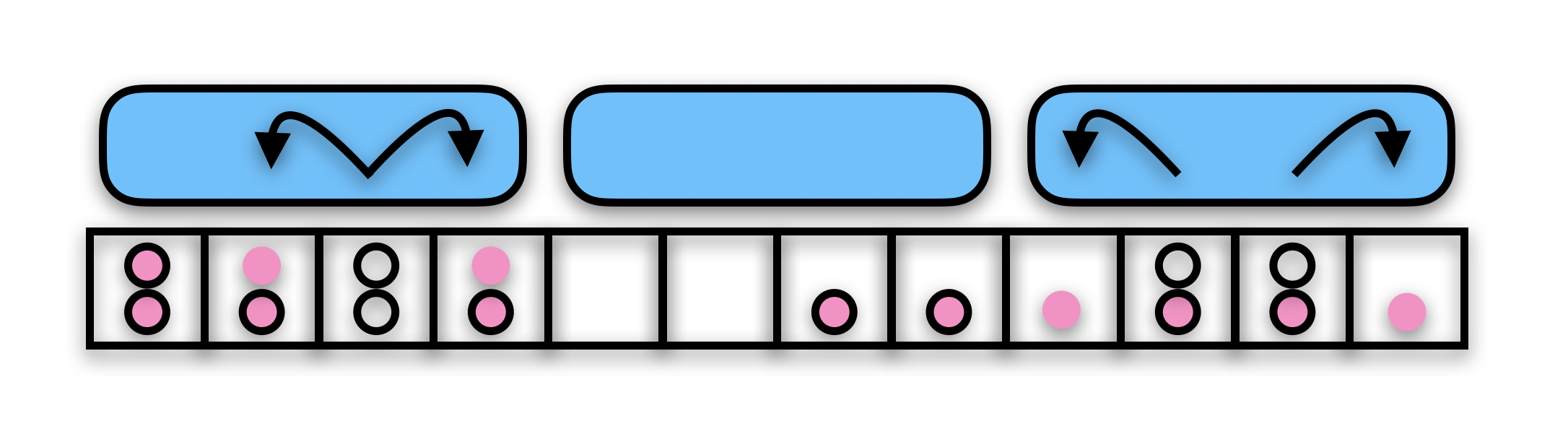}
\caption{\label{fig:markov_circuit} \textbf{The Markov circuit model.} Markov gates (blue) can shuffle the local charge configuration in any way that conserves the total charge and dipole moment of those four sites.  Hollow black circles represent the charge configuration before the action of the gates, and pink circles represent it afterwards. Each site is constrained to have 0, 1, or 2 particles.  The black arrows show some possible actions of the gates; the only transitions that are allowed in this model are ``pair hopping" moves where one (or two) particles hop to the left and one (or two) hop to the right. This model has a detailed balance so that in the long-time steady state all moves are equally likely to occur as the reverse of the same move.}
\end{figure}

We study a model of indistinguishable particles hopping on a one-dimensional lattice of $L$ sites. Interactions are strictly $k$-local and each site $i$ can host $0\leq n_i \le n_\text{max}$ particles. Equivalently, one can consider classical spin models where each site hosts a spin $S = n_{\rm max}/2$ degree of freedom, and particle configurations map to spin configurations in the $z$ basis. Both the total charge (particle number) $N_0=\sum_i n_i$ and dipole moment $N_1 = \sum_i (x_i - x_0) n_i$ are conserved by the dynamics, where $x_0$ can be chosen to be the midpoint of the system, and the sites are at integer positions $1\leq x_i\leq L$. 
The dynamics are given by a kinetically constrained classical Markov circuit. The circuit is made up of layers of $k$-local Markov gates that stochastically map any local charge configuration to any other with the same charge and dipole moment on those $k$ consecutive lattice sites.  A time step consists of one tightly-packed layer of gates with a randomly chosen spatial shift of up to $k-1$ sites. Most of our results are for systems with open boundaries.  Gates that extend past either edge of the system must act only on less than $k$ sites. Fig.~\ref{fig:markov_circuit} depicts one step of the time evolution. We use gates that assign an equal probability to all possible transitions, \textit{i.e.} the ``infinite temperature heat-bath" case. The average of any quantity in the steady state is then given by averaging over all charge configurations in the Krylov sector of the initial state.  Numerically doing this average exactly is only possible for small systems in practice, so we statistically sample the distribution for larger systems for which we can numerically simulate the steady state, while for even larger systems near the phase transition we can only numerically access nonequilibrium dynamics.  

For the rest of this paper we focus on the specific case of $n_\text{max}=2$ and $k=4$, which connects the charge configuration space in exactly the same way as the 4-local spin-1 quantum models of Refs.~\cite{Khemani-Nandkishore2019,Sala-Pollmann2019}, and on systems that have approximately zero total dipole ($N_1 = 0$). Thus the phase diagram we study is as a function of the average charge density $\bar{n}=N_0/L$, which is the parameter that tunes the system from weakly to strongly fragmented. Systems with $n_\text{max}=2$ and $k<4$ do not have a weakly fragmented phase at all, so $k=4$ is the minimal generic case for $n_\text{max}=2$. We provide more explicit details about the $4$-site Markov gates in Appendix~\ref{sec:gate_details}. If one considers systems with a substantially nonzero dipole moment, the equilibrium charge density is spatially nonuniform and the freezing transition occurs locally where the local charge density reaches its critical value. For these systems with zero total dipole, we say the system thermalizes if in the limit of a large system and long times for almost all initial states the time-average of the local density is $\bar n$ at all sites that are far from the ends of the system.  This model has a particle-hole symmetry that dictates that the behavior at $\bar{n}=n$ is the same as that at $\bar{n}=2-n$, so for the remainder of this paper we only consider the regime $\bar{n} \geq 1$.

\section{\label{sec:results} Results}

\subsection{Exact enumeration}

\begin{figure}
\includegraphics[width=1.0\linewidth]{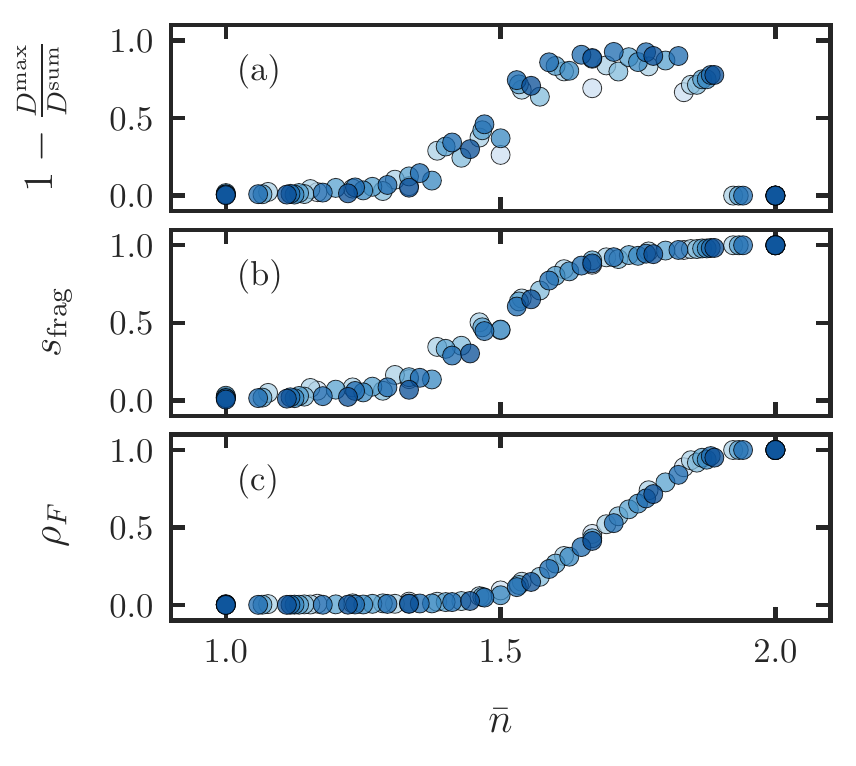}
\caption{\label{fig:small_systems} \textbf{Measures of the fragmentation transition in small systems.} All plots correspond to zero total dipole: $N_1=0$. (a) One minus the fraction of states that belong to the largest Krylov sector. (b) The entropy of fragmentation, as defined in the main text. (c) The infinite-time fraction of frozen sites $\rho_F$ averaged over all possible initial states. All system sizes from $L=12$ (lightest blue) to $L=18$ (darkest blue) sites are represented, but only systems where the total charge and system size have the same parity have states with exactly zero total dipole. These data are symmetric about half filling ($\bar{n}=1$) so only $\bar{n}\ge 1$ are shown.}
\end{figure}

We first study small systems ($L \le 18$) for which we can eliminate finite-time effects, but not finite-size effects, by enumerating all possible configurations and constructing all of the Krylov sectors exactly. This can be done numerically by first constructing a sparse matrix representation of the graph defined by configurations (nodes) and the allowed transitions between them (edges), and then finding the connected components of this graph. This is similar to the approach of Refs.~\cite{Khemani-Nandkishore2019,Sala-Pollmann2019}. The authors of those works defined a quantity to diagnose strong fragmentation that we herein call $D^\text{max} / D^\text{sum}$; it is the ratio of the number of configurations in the largest Krylov sector ($D^\text{max} = \max_j D_j $) to the total number of configurations in all of the Krylov sectors combined ($ D^\text{sum} = \sum_j D_j$). Since charge and dipole are conserved, the index $j$ runs over all Krylov sectors corresponding to a certain total charge $N_0$ and dipole $N_1$ of interest, and $D_j$ is the number of charge configurations in Krylov sector $j$.  To make contact with previous studies we plot this quantity for a range of small system sizes and charge densities in Fig.~\customref{fig:small_systems}{(a)}. This shows an onset of strong fragmentation as the charge density is tuned away from half filling; however finite-size effects are strong and we cannot argue from these data alone that there is a sharp transition in the thermodynamic limit. 

In an attempt to improve on this measure of fragmentation in small systems we define the \textit{entropy of fragmentation}
\begin{equation}
    s_\text{frag} = \frac{- \sum_{j=1}^{j_\text{max}} p_j \log p_j}{\log j_\text{max}},
\end{equation}
where $p_j = D_j / D^\text{sum}$ and $j_\text{max}$ is the total number of Krylov sectors being summed over.  We do this to minimize the effects of $D^\text{sum}$ becoming small as $\bar{n} \to n_\text{max}$ in very small systems; these effects cause the quantity in Fig.~\customref{fig:small_systems}{(a)} to curve downwards as $\bar{n} \to 2$. A weakly fragmented system is characterized by $ \lim_{L\to \infty } s_\text{frag} = 0$, and a strongly fragmented one will have $s_\text{frag}>0$. The entropy of fragmentation is shown in Fig.~\customref{fig:small_systems}{(b)} for comparison. 

Both of the diagnostics of strong fragmentation discussed so far rely on constructing the Krylov sectors exactly, which is prohibitively expensive for all but very small systems. Furthermore the charge density at these small system sizes is not an approximately continuous tuning parameter and this adds difficulty to studying the phase transition. We therefore propose a different diagnostic that can be computed from local measurements and thus can be estimated in larger systems with some finite-time effects without constructing the full Krylov sectors. It is inspired by studies of kinetically constrained classical models of glasses~\cite{Garrahan-vanWijland2007}: 

We propose that the frozen, strongly fragmented phase can be characterized by a nonzero fraction of \textit{frozen} sites in a typical history, where we say that site $i$ is frozen at time $t$ if $n_i$ has remained unchanged for all times $ \le t$.  As time evolves frozen sites can become active, but with this definition active sites never become frozen. This is the same definition of active and frozen sites used in the discussion of ``shielding regions" and the construction of nonthermal states in Refs.~\cite{Khemani-Nandkishore2019,Sala-Pollmann2019}. In those works the focus was on systems at half filling with various ranges of interactions $k$, and on constructing the partially frozen states even in weakly fragmented systems where these are vanishingly rare. To relate the notion of frozen sites back to Krylov sectors and fragmentation: for a given initial state a site will remain frozen for all times if for every configuration in its Krylov sector that site has the same charge $n_i$.  The fraction of frozen sites $\rho_F$ at infinite times, averaged over all initial states, serves as an order parameter heralding the onset of the strongly fragmented phase.  Much of our analysis is focused on this notion of frozen and \textit{active}  regions. In Fig.~\customref{fig:small_systems}{(c)} we show the infinite-time $\rho_F$ averaged over all states with charge density $\bar{n}$ and total dipole $N_1=0$. This shows finite-size indications of the phase transition between weakly and strongly fragmented phases near the nominal critical density $\bar{n}_c = 1.5$, where $\rho_F$ becomes nonzero.

\begin{figure}
\includegraphics[width=1.0\linewidth]{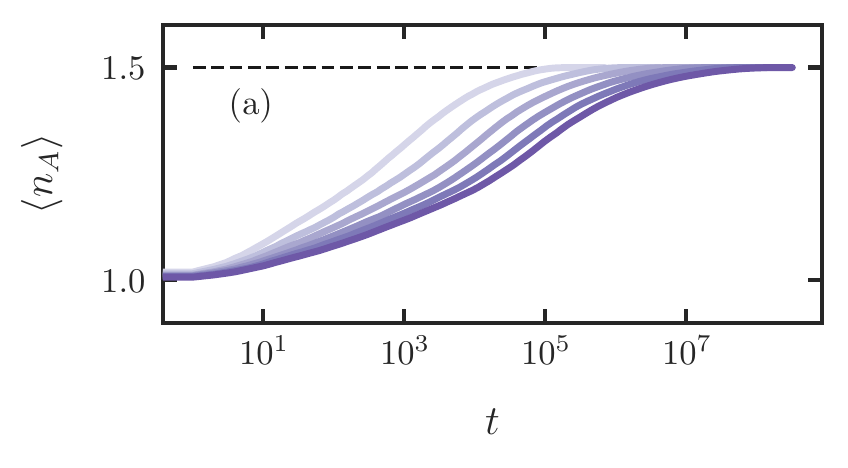}
	\begin{minipage}{0.5\linewidth}
		\includegraphics[width=1.0\linewidth]{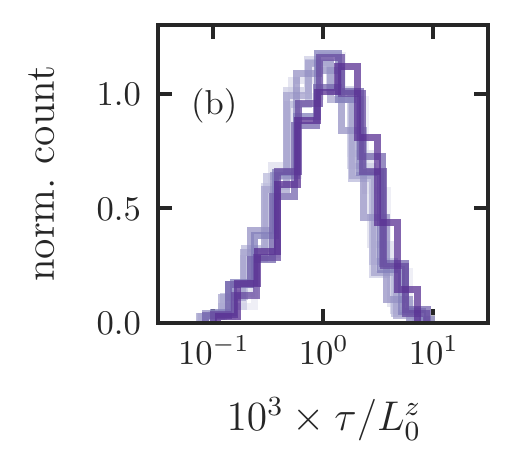}
	\end{minipage}\hfill
	\begin{minipage}{0.5\linewidth}
		\includegraphics[width=1.0\linewidth]{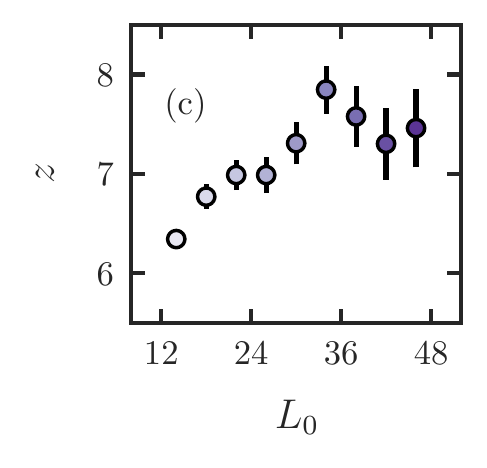}
	\end{minipage}
\caption{\label{fig:active_bubble} \textbf{Ideal active bubble.} (a) The vertical axis is the mean charge density of active sites over all histories and the horizontal axis is the number of time steps (layers in the Markov circuit).  All histories were initialized with a block of length $L_0$ sites with $n_i=1$ embedded in an otherwise frozen infinite system of $n_i=2$. From left to right $L_0=12$ to $32$ in steps of $4$. Each line represents an average over more than $10^3$ histories. (b) Histogram of the scaled time it takes for the active bubble to reach twice its initial size ($n_A=1.5$). The times are scaled by $L_0^z$, where $z=7$. There are at least $10^3$ histories for each curve, and the curves represent $L_0=12$ to $48$ in steps of $4$ (lightest purple to darkest purple). (c) Estimate of the dynamic exponent $z$ as a function of $L_0$ via $z (L_0) = \log (\langle \tau \rangle_+  / \langle \tau \rangle_- ) / \log (L_{+}/L_{-})$, where $L_{\pm} = L_0 \pm 2$ and $\langle \tau \rangle_\pm$ are the corresponding mean values of $\tau$ from the data shown in (b). For example, the first point at $L_0=14$ is obtained by setting $L_- =12$ and $L_+ =16$, then taking the corresponding mean values of $\tau$ from (b) and calling those mean values $\langle \tau \rangle_-$ and $\langle \tau \rangle_+$. Error bars represent the standard error.}
\end{figure}

\subsection{Active vs. frozen regions}

In the vicinity of the critical density large systems typically contain regions that are initially frozen and regions that are initially active. The regions that are initially frozen are typically contiguous blocks of sites with a local charge density near $n_\text{max}=2$, and the initially active regions are typically contiguous blocks of sites where the charge density is closer to half filling and below the critical density. For example, by identifying the initially frozen sites as the sites that cannot evolve to any other value of charge during the first step of time evolution, no matter what spatial shift that first layer of the Markov circuit has, we find that systems with $L=10^6$ and $\bar{n}=1.5$ contain contiguous blocks of frozen ($F$) sites with mean charge density $\langle n_F \rangle \cong 1.95$ and active ($A$) blocks with $\langle n_A \rangle \cong 1.28$, where the averages are taken over equally weighted blocks. As time goes on, a frozen site can become active if a neighboring active block is a good enough charge bath to move some of its charge, so the active blocks can thus grow and merge.  As the active blocks grow and occupy a larger fraction of the system their charge density increases. The growth stops when the charge density of the active block approaches the critical value and the active block thus stops being able to move the charges on the neighboring frozen sites. 

\subsection{Single active blocks \label{subsec:single_active_blocks}}

In order to understand this process we study the idealized scenario of a half-filled initial block of $L_0$ contiguous sites, with initial charges $n_i=1$ on all of those $L_0$ sites, embedded in an otherwise fully-filled system of infinite length with initial charges $n_i =2$ on all other sites. This is in some ways similar to the idealized scenarios used to study the effects of thermal inclusions in an MBL system and their relevance to the MBL transition~\cite{DeRoeck-Huveneers2017,Luitz-DeRoeck2017,Thiery-DeRoeck2018,Dumitrescu-Vasseur2019,Gopalakrishnan-Huse2019,Morningstar-Huse2019}. We stochastically evolve this specific initial configuration multiple times, \textit{i.e.}, we run multiple histories.  As a history evolves we keep track of which sites have become active and the corresponding charge density $n_A$ of the growing active bubble in that history.  When the length of the block of active sites has grown to $L_A$, then the charge density of the active block is $n_A=2-(L_0/L_A)$.  In Fig.~\customref{fig:active_bubble}{(a)} we show the resulting time evolution of the history-averaged $n_A$ for active bubbles of initial size $L_0=12$ to $32$.  In each such history with even $L_0\leq 48$ the active bubble converges to exactly twice its initial length and a charge density of $n_A=1.5\cong\bar{n}_c$, where it finally ceases to be a good enough charge bath to move charges from the neighboring frozen sites. We have been able to show by an explicit iterative construction that even length bubbles can grow to twice their length, thus showing that $\bar{n}_c\geq 1.50$; we suspect that the exact value of the critical density does saturate this bound, but we have not yet been able to prove that.

We have also studied initial blocks of $(L_0+1)$ contiguous sites where $L_0$ of the sites in the block are singly occupied, while one of the sites within the block and all sites outside of the block are doubly occupied.  Such active bubbles all appear to stop their growth with length $2L_0$ contiguous active sites, with the sole exceptions of all cases where the center of mass of the active bubble is precisely on a site.  These latter cases instead stop with length $(2L_0-1)$.  We have numerically examined many bubbles with even and odd $L_0\leq 48$ and various locations of their centers of mass, not finding any exceptions to this behavior.

We denote the time at which the active bubble in any particular history first reaches its final length by $\tau$ and extract a dynamical scaling exponent $z \cong 7.0(5)$, such that the history-averaged $\tau$ scales as $\langle\tau\rangle \propto L_0^z$, by collapsing the distribution of $\tau$ over histories for each system size, as shown in Fig.~\customref{fig:active_bubble}{(b)}. However the range of $L_0$ over which we were able to compute $\langle \tau \rangle$ is quite small (on a $\log$ scale), and statistically significant drift towards higher $z$ is present (see Fig.~\customref{fig:active_bubble}{(c)}), thus it is possible that $z$ drifts upwards indefinitely with increasing $L_0$, or it may converge to a finite value. We therefore leave our estimate of $z\cong 7$ as a lower bound. We also performed the same ``ideal active bubble" experiments but with active blocks that were initialized in a random configuration at half filling embedded in a fully-filled system. In that case the density of the active blocks of almost all samples still converges to $n_A=1.5$, and the fraction of samples for which this is not true decreases with $L_0$. We therefore expect that the ideal scenario studied above is representative of large active bubbles embedded in typical large systems. In a typical system there will be many such active blocks, and they will expand and merge as time goes on. We propose that these blocks expand until they self-tune to the critical charge density where they are no longer good charge baths, and the critical point is where the system is just able to be completely covered by the largest, slowest active block at the latest times. Just above the critical density, a finite density of frozen sites will persist indefinitely and block charge transport, preventing thermalization. 

\subsection{Approximate model for the transition}

By considering isolated active bubbles embedded in otherwise frozen systems we have developed the idea that an active bubble typically unfreezes the immediately adjacent frozen sites and, in doing so, increases its average charge density and length.  This process continues until the critical charge density is reached, at which point the growth stops because the active bubble is no longer an effective charge bath. We therefore propose an approximate scheme for partitioning a system with $\bar n > \bar n_c$ into frozen and active blocks at time $t = \infty$: Given a system at $t=0$ we first compute its initially active ($A$) and frozen ($F$) blocks of sites by considering which sites \textit{could} evolve to a different charge during a hypothetical first
%time step
layer of Markov gates, regardless of the spatial shift of that first layer.
%of gates in the Markov circuit.
Once we have set up the blocks of sites labelled as initially active, we start to loop over all of these active blocks, one by one, to allow them to expand if possible. An active block is allowed to expand by one site to the left if incorporating that left-neighboring site (and its charge) does not push the average charge density of that active block over the nominal critical charge density ($n_A \le \bar{n}_c$). The same condition is used for expanding by one site to the right. If an active block qualifies to expand in only one direction, it does so. If it qualifies to expand in both directions, a direction is randomly chosen, and the expansion is carried out. We iterate this procedure, allowing a single active block to expand by one site at a time as long as it does not overlap with another active block and maintains $n_A \le \bar{n}_c$ as described above, before moving on to the next active block in the loop. Note that in this approximate model we only keep track of the average charge density of active blocks, assuming the system is thermalizing within these blocks, and the charge configuration on the frozen sites remains as it was initially, so we do not do explicit updates of the charge configuration as in the exact dynamics. 
%We then loop over all of the active blocks, allowing each block to expand by incorporating the sites and the charges in neighboring frozen blocks.  This expansion proceeds in either or both directions and is continued as long as the active block's average charge density satisfies $n_A < \bar{n}_c$ and it does not overlap with another active block. 
After allowing all of the active blocks to grow like this, any two active blocks that have come into direct contact are merged and considered as one block from then on. We iterate this process of looping over active blocks allowing them to grow as much as possible, then merging ones that have come in contact, until it has converged, \textit{i.e.} all active blocks reach density $n_A=\bar{n}_c$ and are thus unable to grow farther.  This approximate model of the steady state must be given the critical density as an input parameter, so we choose to use $\bar{n}_c=3/2$, which is our current best (and perhaps exact) estimate of the critical point. For $\bar n \leq \bar n_c$ and typical initial states of large systems, this model grows one active block that encompasses the entire system and thus thermalizes.

\begin{figure}
\includegraphics[width=1.0\linewidth]{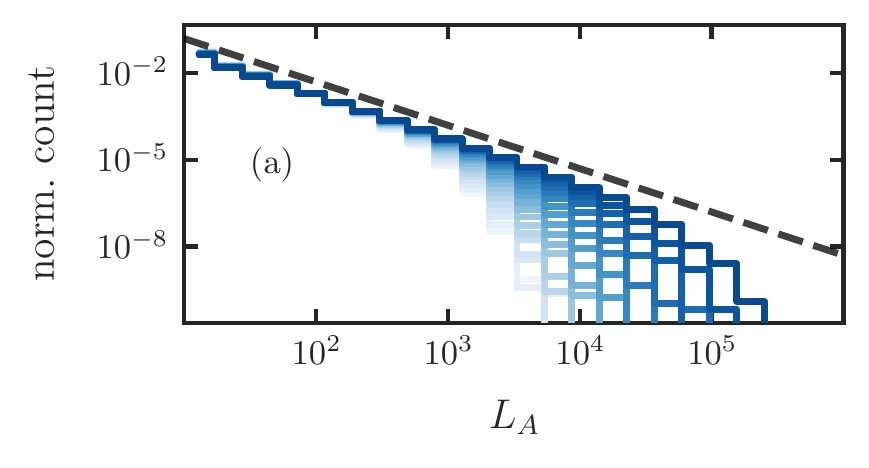}
\begin{minipage}{0.5\linewidth}
	\includegraphics[width=1.0\linewidth]{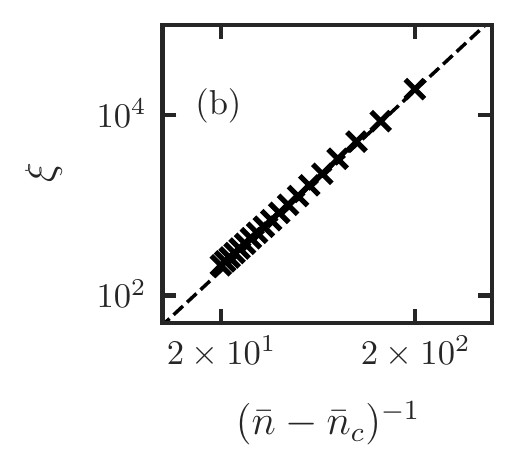}
\end{minipage}\hfill
\begin{minipage}{0.5\linewidth}
	\includegraphics[width=1.0\linewidth]{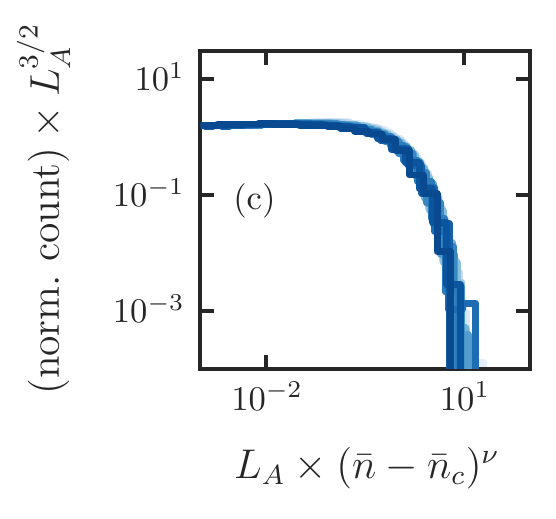}
\end{minipage}
\caption{\label{fig:approx_scheme} \textbf{Approximate model of the steady state.} (a) Distributions (``normalized counts'') of the final lengths of the active blocks. Charge densities go from $\bar{n}=1.505$ (darkest blue) to $1.550$ (lightest blue) in steps of $2.5\times 10^{-3}$. Each line represents $10^2$ samples of length $L=10^6$ drawn randomly from all configurations with the correct amount of total charge. The dashed line is $\sim L_A^{-\alpha}$ with $\alpha=3/2$, as is expected for the small $L_A$ regime. (b) The correlation length $\xi=\langle L_A^2 \rangle / \langle L_A \rangle$. The dashed line represents the expected scaling $\xi \propto (\bar{n}-\bar{n}_c)^{-\nu}$ where $\nu=2$. (c) Scaled versions of the curves in part (a). }
\end{figure}

Although we will provide analytic arguments for all of the following results, it is convenient to first present the numerical data from simulations of this approximate model of the steady state. At each of a series of charge densities $\bar n$ approaching $\bar{n}_c$ from above we simulate this process on $10^2$ samples of large systems ($L=10^6$), initialized randomly from all configurations with precisely the correct amount of total charge, and store the distribution of the lengths $L_A$ of the final active blocks; the final {\it frozen} blocks remain of length of order one for $\bar n$ near but above $\bar n_c$. The resulting data are shown in Fig.~\customref{fig:approx_scheme}{(a)}. The distributions, which we denote by $p_{\bar{n}} (L_A)$, approach a power law $\sim L_A^{-\alpha}$ with $\alpha=3/2$ as $\bar{n}$ approaches $\bar{n}_c$ from above. 

We define a correlation length $\xi = \langle L_A^2 \rangle / \langle L_A \rangle$, which we compute from the data at each value of $\bar{n}$, and we plot this in Fig.~\customref{fig:approx_scheme}{(b)}. These averages are over all active blocks, with each block equally weighted independent of its length.  Defined in this way, $\xi$ is a characteristic active block length where the distribution changes from a power law for $L_A\ll\xi$ to something steeper for $L_A\gg\xi$.  From this we see that $\xi$ diverges at the critical density as $\xi \propto (\bar{n}-\bar{n}_c)^{-\nu}$, with $\nu=2$ being consistent with the data. We assume the scaling ansatz $p_{\bar{n}} (L_A) \propto L_A^{-\alpha} f(L_A / \xi)$, where $f$ is a universal function, to collapse the data of Fig.~\customref{fig:approx_scheme}{(a)} and we show the scaled distributions in Fig.~\customref{fig:approx_scheme}{(c)}. With this scaling, $\langle L_A^p\rangle\sim\xi^{(p+1-\alpha)}$ for any power $p>\alpha-1$, which is why we defined $\xi$ the way we did, rather than as $\langle L_A\rangle$, for example.

To summarize: This approximation yields the predictions that the critical point is characterized by a power-law distribution of the lengths of active blocks with an exponent of $\alpha = 3/2$, and by a correlation length $\xi$ which diverges with the exponent $\nu=2$ as the critical point is approached from the strongly fragmented phase.

In order to argue analytically for the critical power-law distribution $p_{\bar{n}_c} (L_A) \propto L_A^{-3/2}$ we consider a critical ($\bar n = \bar n_c$) system with an embedded active block whose total charge is initially $\Delta N_{0,A}$ below what it would be if the block were at the critical density. The block will begin to grow and each time its length increases by $\Delta L_A$ it gains an amount of charge $\Delta L_A \bar{n}_c + \delta$, where $\delta$ is a random variable with mean zero. Therefore $\Delta N_{0,A}$, the total charge deficit of the block relative to the critical total charge, does an unbiased random walk in one dimension with an ``absorbing wall" at $\Delta N_{0,A}=0$ where the active block becomes critical and can no longer grow. The analogous real-space problem is that of a random walk $x(t)$, where $x$ is position and $t$ is time, that begins at a negative position $x(t_0)=-x_0$ and we want to know: What is the distribution of final times $t_f$ at which the walker reaches $x=0$ for the first time? This can be solved in the continuum by the method of images, with the result being that the distribution of final times goes like $p(t_f)\sim t_f^{-3/2}$, which is analogous to the observed result that $p_{\bar{n}_c} (L_A) \sim L_A^{-\alpha}$ with $\alpha = 3/2$.

We can also argue for the result $\nu=2$: The frozen blocks of the system have an average charge density $\langle n_F\rangle$ that is always well above $\bar n_c$ and close to $n_\text{max}$, and we have argued that large active blocks will converge to the critical charge density $n_A=\bar{n}_c$ at late times. Therefore we consider an approximation where all frozen blocks have $n_F = n_\text{max} = 2$ and all active blocks have $n_A = \bar{n}_c$ as $t\to \infty$. In that case the shifted charge density can be expressed as $\bar{n}-\bar{n}_c =(n_\text{max} -\bar{n}_c ) \langle L_F \rangle / (\langle L_A \rangle + \langle L_F \rangle)$, and since $\langle L_F \rangle$ does not diverge at the transition this implies $\langle L_A \rangle \propto (\bar{n}-\bar{n}_c)^{-1}$ as the critical point is approached within the frozen phase. If the distribution of $L_A$ takes the form $p_{\bar{n}} \propto L_A^{-3/2} f(L_A / \xi)$, then the identity $\langle L_A \rangle = \int \ell p_{\bar{n}}(\ell) d\ell$ and dimensional analysis implies $\langle L_A \rangle \propto \sqrt{\xi}$ and therefore $\xi \propto (\bar{n}-\bar{n}_c)^{-2}$, \textit{i.e.} $\nu=2$. We can also obtain the critical exponent $\beta$ of the order parameter $\rho_F \sim (\bar{n}-\bar{n}_c)^{\beta}$ (the fraction of frozen sites) via similar reasoning: Again because $\langle L_F \rangle$ does not diverge at the transition and $\langle L_A \rangle$ diverges as $(\bar{n}-\bar{n}_c)^{-1}$, the identity $\rho_F = \langle L_F \rangle / (\langle L_F \rangle + \langle L_A \rangle)$ implies $\beta=1$.

\begin{figure}
\includegraphics[width=1.0\linewidth]{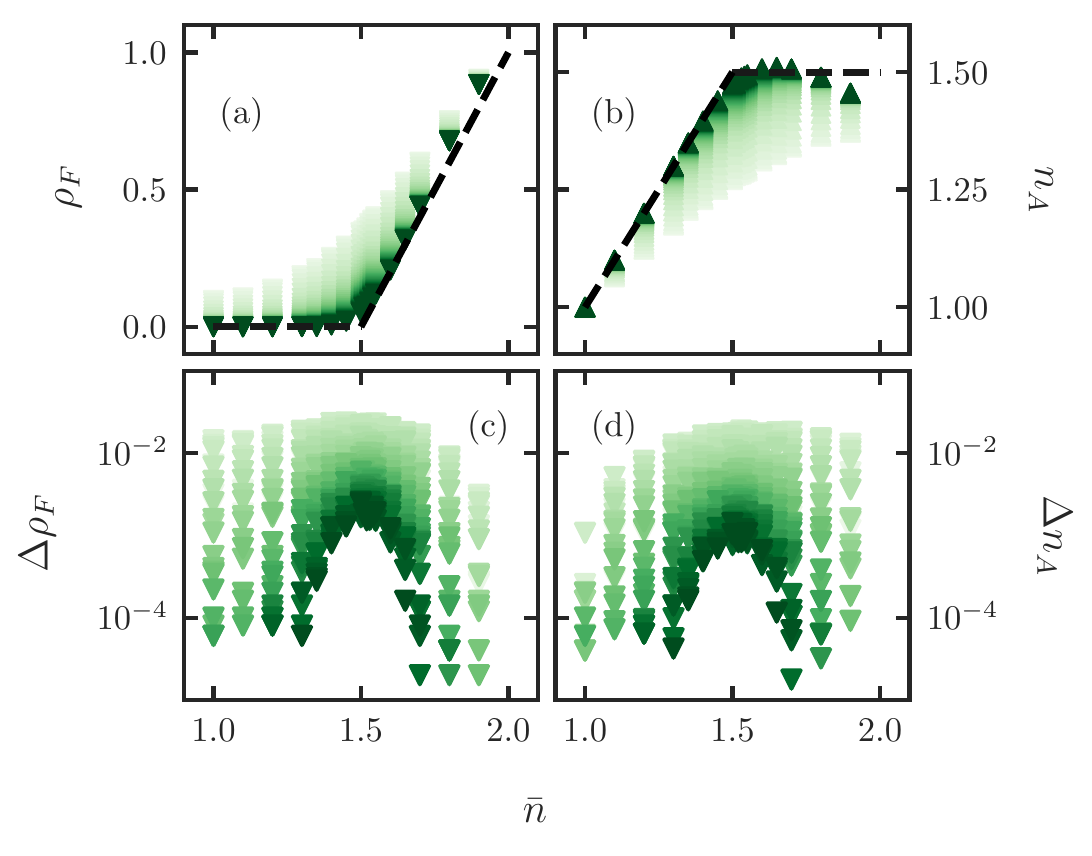}
\caption{\label{fig:mc_rhoF_nA} \textbf{Time evolution of typical large systems.} All plots represent the same simulations of the full Markov circuit dynamics of systems with $L=10^4$ out to times $t=10^8$. Each data point is averaged over $5-10$ samples. Quantities are shown at $40$ times from $t=0$ to $10^8$ spaced evenly on a logarithmic scale (light to dark green). (a) The fraction of frozen sites. The dashed curve represents the simplified theoretical prediction detailed in the main text. (b) The charge density of the total area in the system that has been labelled active, \textit{i.e.} $n_A = \sum_{i \in A} n_i / (1-\rho_F)L$. (c) and (d) The absolute difference in $\rho_F$ and $n_A$, respectively, from each of the $40$ times at which $\rho_F$ and $n_A$ are shown to the next.}
\end{figure}

\subsection{Simulation results}

At this point our understanding has been built up from idealized initial conditions and simplified models. We now go back to the full dynamics of large systems in an effort to validate as much of the picture we developed above as possible. We randomly initialize systems with definite charge density $\bar{n}$ of size $L=10^4$ and stochastically sample their full dynamics up to $t=10^8$ layers of the Markov circuit. This is still very early times for such large systems (see Fig.~\ref{fig:active_bubble}), so we are targeting the infinite system, finite time limit, contrary to the infinite time, small $L$ results shown in Fig.~\ref{fig:small_systems}. In Fig.~\ref{fig:mc_rhoF_nA} we show the time evolution of the fraction of frozen sites $\rho_F$ and the charge density of the active fraction of the system $n_A$. Fig.~\customref{fig:mc_rhoF_nA}{(a)} is consistent with the result of Fig.~\customref{fig:small_systems}{(c)} for small systems and it appears that $\rho_F \propto (\bar{n}-\bar{n}_c)^\beta$ with $\beta=1$ and $\bar{n}_c = 1.50$ is consistent in the strongly fragmented phase, as our simplified theory and previous numerical results suggested. The slight deviation of the late-time data from the simplified theoretical prediction (dashed line) deep in the frozen phase is understood: In typical systems the frozen blocks contain some sites that do not have charge $n_{i}=n_\text{max}$, thus $\rho_F \to 1$ more rapidly than $\bar{n}\to 2$, and indeed this was observed in small systems as well (see Fig.~\ref{fig:small_systems}). This effect is significant deep in the frozen phase because that is where the frozen blocks are largest and can contain a small number of isolated frozen sites that have $n_i<n_\text{max}$.
 
In the second plot, Fig.~\customref{fig:mc_rhoF_nA}{(b)}, we show that the charge density of the combined active blocks of the system is converging to $n_A=\bar{n}$ from below in the active phase (low density active blocks activate high density frozen blocks until the system is entirely active) and in the frozen phase near the critical point there is an extended regime where the active portion of the system self-tunes to the critical density $n_A \to \bar{n}_c$, as we have suggested. Deep in the frozen phase almost all of the active bubbles are very small and many of them have a two-state Krylov sector (...2220222... and ...2212122...) that grows only to density $n_A=4/3$, so the average $n_A$ becomes noticeably less than $\bar{n}_c$ (dashed line), as is visible in Fig.~\customref{fig:mc_rhoF_nA}{(b)}. Finally in Fig.~\customref{fig:mc_rhoF_nA}{(c),(d)} we show how $\rho_F$ and $n_A$ are converging to their final values at each $\bar{n}$ in order to demonstrate that $\bar{n}_c=1.5$ is a reasonable estimate of the critical density for typical large systems. Since we are only able to access rather early time dynamics, we have not been able to test the prediction that the critical distribution of the lengths of active blocks goes like $p_{\bar{n}_c}(L_A)\propto L_A^{-3/2}$. This may be because we are not able to build up large active blocks from small ones during the full simulation of large typical systems during the accessible times. Thus we leave the results presented in Fig.~\ref{fig:approx_scheme} as predictions of our simplified theory, not yet tested by simulations of the exact dynamics.

\subsection{Subdiffusion in the thermalizing phase and charge autocorrelations at criticality}

Most of our analysis so far has focused on the properties and statistics of active bubbles embedded in frozen or near-critical systems. In this subsection we study the dynamics in the thermalizing phase and at the critical point by examining the averaged charge-charge correlation function $C(x_i,t)=\langle (n_i(t) - \bar{n})(n_0(0) - \bar{n}) \rangle$ in systems with $L=10^3$ sites. The results presented here are the only ones for which we employ periodic boundaries because it allows for averaging over all sites.

\begin{figure}
\begin{minipage}{0.5\linewidth}
	\includegraphics[width=1.0\linewidth]{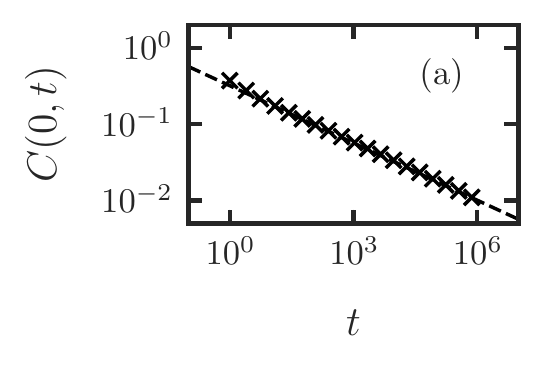}
\end{minipage}\hfill
\begin{minipage}{0.5\linewidth}
	\includegraphics[width=1.0\linewidth]{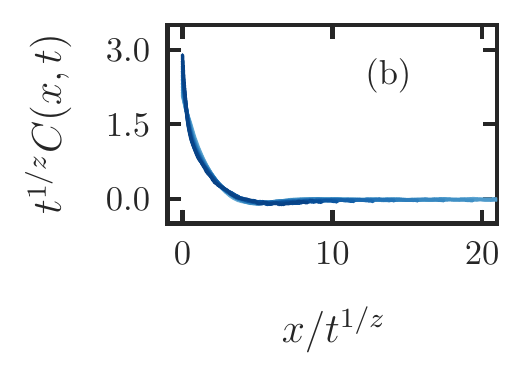}
\end{minipage}
\begin{minipage}{0.5\linewidth}
	\includegraphics[width=1.0\linewidth]{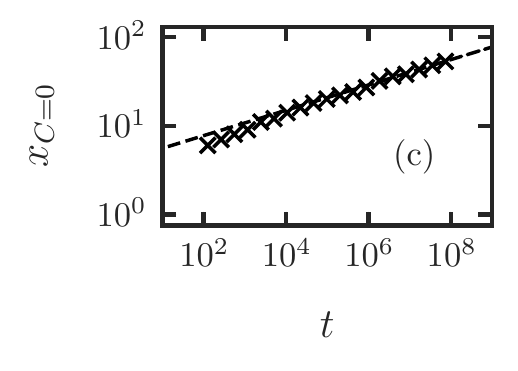}
\end{minipage}\hfill
\begin{minipage}{0.5\linewidth}
	\includegraphics[width=1.0\linewidth]{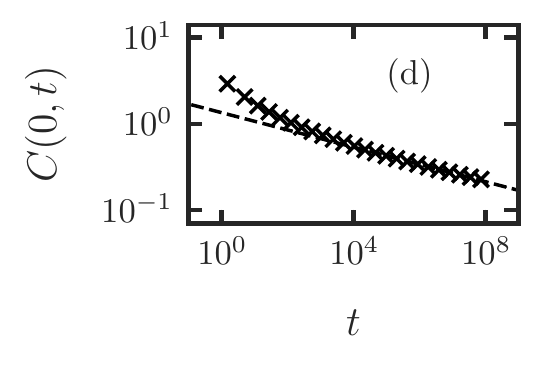}
\end{minipage}
\caption{\label{fig:autocorr} \textbf{Autocorrelations of charge density fluctuations in half-filled and critical systems.} (a) On-site charge autocorrelations at half filling $\bar{n}=1$. The dashed line represents the subdiffusive scaling $C(0,t) \propto t^{-1/4}$. Data for $C(0,t)$ (black crosses) was generated by simulating the dynamics of $10^3$ samples of systems with $L=10^3$ sites and periodic boundaries out to time $t=10^6$. Averaging was done over all sites, samples, and windows of time surrounding each data point shown above. (b) The scaled charge autocorrelation function for critical systems plotted as a function of $x$ for several times $10^4 < t < 10^8$. The scaling exponent was taken to be $z=7$. Data (black crosses) was generated by simulating the dynamics of $8\times 10^2$ samples with $L=10^3$ sites and periodic boundaries out to time $t=10^8$. Averaging was done in the same way as in part (a). The curves are associated with times spaced evenly on a log scale (light to dark blue). (c) The distance at which charge autocorrelations crossover from positive to negative, $\textit{i.e.}\ C(x_{C=0},t)=0$, in critical systems ($\bar{n}=1.5$). The dashed line represents the scaling $x_{C=0} \propto t^{1/z}$ again with $z=7$. The data is the same as for part (b). (d) On-site charge autocorrelations at the critical density $\bar{n}=1.5$. Data and averaging is the same as in parts (b) and (c). The dashed line represents the scaling $C(0,t)\propto t^{-1/10}$.}
\end{figure}

In Fig.~\customref{fig:autocorr}{(a)} we show the averaged on-site charge autocorrelation function $C(0,t)$ at charge density $\bar{n}=1$ in the middle of the thermalizing phase. We find that the charge dynamics of the weakly fragmented phase is subdiffusive with a characteristic space-time ($x$-$t$) scaling $x \propto t^{1/4}$, and we believe this dynamical scaling is characteristic of dipole-conserving systems that thermalize. 

An argument for the subdiffusive $x \propto t^{1/4}$ scaling in the thermalizing phase goes as follows: We work with a one-component hydrodynamics of the charge density $n(x,t)$, which is now a continuous-valued field in the coarse-grained system. Since the dipole moment is conserved, the fundamental process for moving charge is not one that displaces charge from position $x$ to $x+dx$, instead it is one that displaces equal amounts of charge from $x$ to $x+dx$ and $x-dx$, and thus preserves the dipole moment. We call the density of these processes the pair hopping density $p(x,t)$, and take the convention that a positive $p(x,t)$ depletes the charge at position $x$. Since these processes are isotropic, a gradient of them can generate a local charge current $j$, $\textit{i.e.}$ $j \propto - \nabla p$. Assuming the system is entropy-driven and that the entropy density $s$ at position $x$ only depends on the density at that position, the change in entropy due to local pair hopping is $\propto -\nabla^2 n$ when expanding around a uniform-density equilibrium. In order to understand where this comes from we can again imagine the displacement of an amount of charge $\delta n$ from site $x$ to sites $x+dx$ and $x-dx$ in equal amounts. To leading order the entropy change due to this would be 
\begin{align}\label{eqn:ds}
    \delta s \cong s'[n(x-dx)] \frac{\delta n}{2} - s'[n(x)] \delta n + s'[n(x+dx)] \frac{\delta n}{2},
\end{align}
where the prime denotes a derivative with respect to the argument. Near the maximum-entropy equilibrium $n(x)=\bar{n}$ we have $s'[n] \cong s'[\bar{n}]+ s''[\bar{n}](n - \bar{n})$ and $n(x \pm dx) \cong n(x) \pm \nabla n(x) dx + (1/2) \nabla^2 n(x) dx^2$. Plugging these into Eqn.~(\ref{eqn:ds}) we get that the leading entropy change is $\propto -\nabla^2 n$ as stated. This implies the pair-hopping density is driven by $p \propto -\nabla^2 n $. Thus charge currents are driven by the ``Fick's Law" $j \propto \nabla^3 n$. Combining this with charge conservation $\dot{n} +\nabla j = 0$ yields the desired result $\dot{n} \propto - \nabla^4 n$, which implies an $x \propto t^{1/4}$ scaling for the relaxation of charge density perturbations. This subdiffusive scaling is also in agreement with a more complete theoretical framework for the hydrodynamics of thermalizing systems of fractons that was recently developed~\cite{Gromov-Nandkishore2020}, and it was also recently observed in a cold atom quantum simulation of a ``tilted" Fermi Hubbard model~\cite{Guardado-Sanchez-Bakr2019} with an emergent dipole moment conservation on lengthscales larger than the lattice spacing. Such a platform could provide an experimental testing ground for studying weak and strong Hilbert space fragmentation in quantum systems, and for investigating phase transitions between the two like the one studied in this paper.

Finally, in Fig.~\customref{fig:autocorr}{(b)-(d)} we show the charge-charge correlations of systems exactly at the critical density $\bar{n}=1.5$. A scenario supported by our data is as follows: At late times and long distances the charge autocorrelation function at the critical density scales as $C(x,t) = t^{-1/z} F(x/t^{1/z})$ (see Fig.~\customref{fig:autocorr}{(b)}), where $F$ is a scaling function and $z\cong 7$, consistent with our earlier results shown in Fig.~\ref{fig:active_bubble}. Support for $z \cong 7$ can be found by tracking $x_{C=0}$, the relative position at which correlations cross over from positive to negative, as a function of time. This quantity is shown in Fig.~\customref{fig:autocorr}{(c)} to scale as $x_{C=0}\propto t^{1/7}$. At small $x$ the charge autocorrelations decay more slowly, for example $C(0,t)\propto t^{-1/10}$ (see Fig.~\customref{fig:autocorr}{(d)}). This slower decay at small distances is consistent with a power law divergence of the scaling function $F$ as its argument goes to zero: If $F(y)\propto y^{-\eta}$ at small $x$ then $C(x,t) \propto t^{-(1-\eta)/z}$ at those small distances. Thus our data indicates an exponent $\eta \cong 0.3$. We believe this divergence at short distances is caused by the low density of small frozen blocks of sites that remain at finite times; this density was also found to decay as roughly $t^{-1/10}$ in critical systems.

\section{\label{sec:summary} Summary and outlook}

We studied the freezing phase transition encountered when tuning the average charge density of a one-dimensional system that conserves its total charge and dipole moment. In the thermalizing phase the system is weakly fragmented in the sense that almost all charge configurations are connected to each other by the stochastic dynamics of our model, and thus the system eventually reaches equilibrium from typical initial states. In the frozen phase the set of global charge configurations is strongly fragmented, so the set of charge configurations shatters into exponentially many dynamically disconnected sectors, with no single sector being dominant over all others, and we explained how this manifests itself in terms of the growth and isolation of locally weakly fragmented active bubbles. These bubbles act as charge baths and help to thermalize the initially frozen regions of the system, but in the strongly fragmented phase active bubbles eventually stop growing and they remain isolated from each other for all time. We studied the dynamic scaling of these active regions and found that they self-tune to the critical charge density near the critical point. Based on this understanding we developed a solvable simplified model that yields several predictions about the static properties of the critical point. We also showed that in the weakly fragmented phase the charge dynamics is subdiffusive, and discussed the prospects for studying these dipole-conserving systems experimentally.

This work has focused on the classical aspects of Hilbert space fragmentation in dipole-conserving systems. Two related question for future research are then:  What are the distinctly quantum aspects of systems that conserve dipole moment? and: How is the physics of dipole-conserving systems related to the glassy phenomenology of other classical kinetically constrained models?

$\textit{Note added:}$ While finalizing this manuscript we became aware of related work by J. Feldmeier, P. Sala, G. de Tomasi, F. Pollmann, and M. Knap~\cite{Feldmeier-Knap2019}. Where our results overlap they agree. 

\begin{acknowledgments}

We would like to thank Joel Lebowitz and Sanjay Moudgalya for helpful discussions. A.M. acknowledges the support of the Natural Sciences and Engineering Research Council of Canada (NSERC).  D.A.H. was supported in part by a Simons Fellowship and by DOE grant DE-SC0016244.

\end{acknowledgments}

\appendix

\section{$4$-site Markov gates \label{sec:gate_details}}

The gates used in this work act on four sites, each with charge $n_i \in \{ 0,1,2 \}$. There are $3^4 = 81$ different local charge configurations ($n_1 n_2 n_3 n_4$) that a gate can encounter as an input. When a gate is applied to any of these inputs, the output is a randomly chosen (with equal probabilities) configuration on those four sites with the same total charge and dipole moment as the input. Thus the action of the gate can be explicitly specified by listing the groups of configurations on four sites with the same charge and dipole moment. There are $26$ configurations that can only map to themselves under the action of the gate, $28$ that can map to themselves or one other configuration ($14$ groups of $2$), and $27$ that can map to themselves or two other configurations ($9$ groups of $3$). All of these groups are labelled by their own distinct $(N_0, N_1)$ pair. These groups (and thus the allowed local transitions) are shown in Table~\ref{tab:gate_sectors}.

\begin{table}
\begin{tabular}{ |c c c| }
    \hline
    0120 & 0201 & 1011 \\
    \hline
    0121 & 0202 & 1012 \\
    \hline
    0210 & 1020 & 1101 \\
    \hline
    0211 & 1021 & 1102 \\
    \hline
    0220 & 1111 & 2002 \\
    \hline
    1120 & 1201 & 2011 \\
    \hline
    1121 & 1202 & 2012 \\
    \hline
    1210 & 2020 & 2101 \\
    \hline
    1211 & 2021 & 2102 \\ 
    \hline
\end{tabular}
\hspace{0.5in}
\begin{tabular}{ |c c| }
    \hline
    0020 & 0101 \\
    \hline
    0021 & 0102 \\
    \hline
    0110 & 1001 \\
    \hline
    0111 & 1002 \\
    \hline
    0200 & 1010 \\
    \hline
    0212 & 1022 \\
    \hline
    0221 & 1112 \\
    \hline
    1110 & 2001 \\
    \hline
    1200 & 2010 \\
    \hline
    1212 & 2022 \\
    \hline
    1220 & 2111 \\
    \hline
    1221 & 2112 \\
    \hline
    2120 & 2201 \\
    \hline
    2121 & 2202 \\
    \hline
\end{tabular}
\caption{\label{tab:gate_sectors} \textbf{Allowed transitions for the 4-site Markov gate.} Configurations $n_1 n_2 n_3 n_4$ that have the same charge and dipole moment are shown in groups of three (left) or two (right). Only transitions between states within each group are allowed. The two sets of groups shown contain a total of 55 configurations; the other 26 (of 81) belong to single-state groups, so they are not shown.}
\end{table}

\section{Super-critical active bubbles \label{sec:super_critical}}

\begin{figure}
\includegraphics[width=1.0\linewidth]{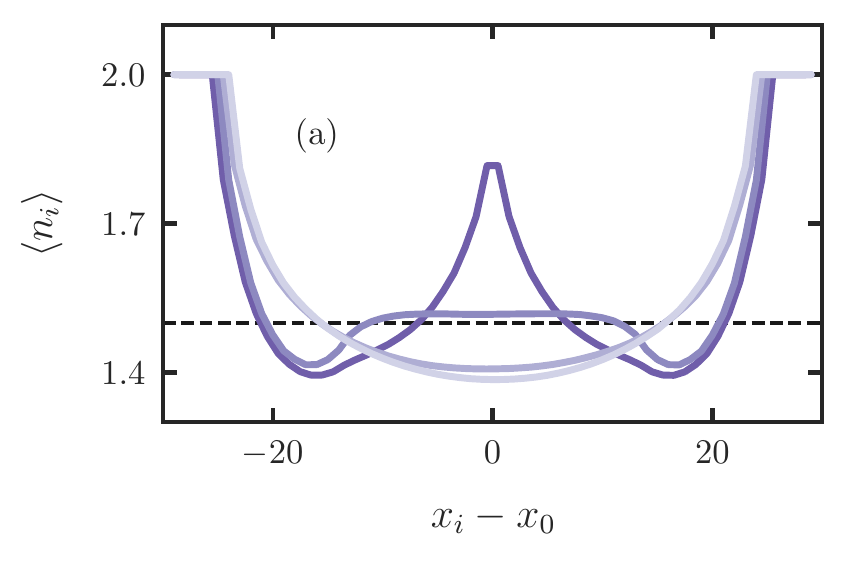}
\includegraphics[width=1.0\linewidth]{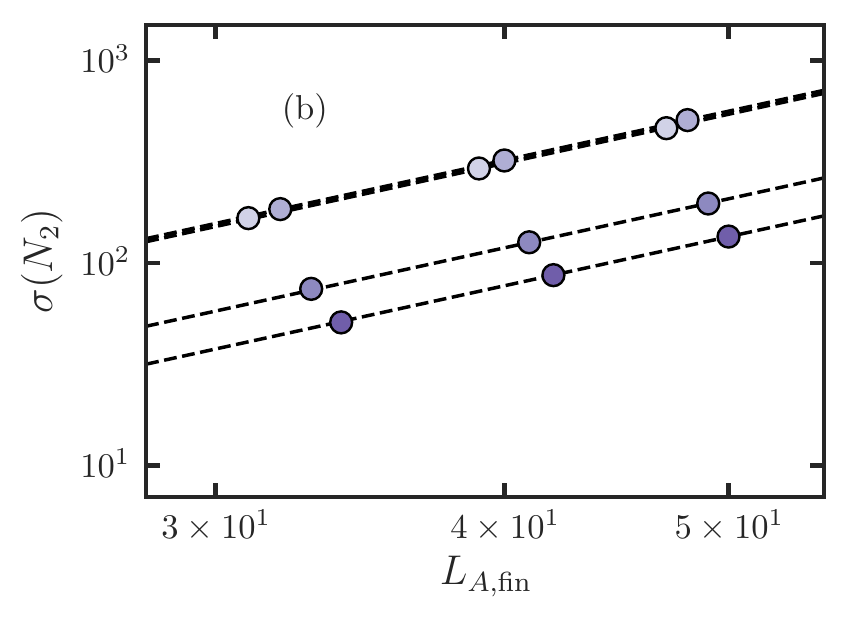}
\caption{\label{fig:double_bubble} \textbf{Equilibrium of two merged active blocks.} (a) Equilibrium charge distributions. The dashed line is at $\langle n_i \rangle = 1.5$, the critical density. The purple curves correspond to the scenarios of two initial blocks of length $L_A=12$ sites with $n_i=1$ separated by an initial block of length $L_A+\{-1,0,1,2\}$ sites (light to dark) with $n_i=2$. Each system was time evolved for $t_f=10^{10}$ time steps and the curves represent the steady-state charge density distributions over $4$ samples during the final $2.5\times 10^9$ time steps. (b) The standard deviation of the quadrupole moment fluctuations of the final active sites in equilibrium. The four right-most data points correspond to the same simulations that generated the curves in part (a), with $L_A=12$. The center and left groups of four data points correspond to $L_A=10$ and $L_A=8$, respectively. The lines are fits of the form $\sigma(N_2) = A L_{A,\text{fin}}^{5/2}$, with $A=3\times 10^{-2},\ 3\times 10^{-2},\ 1\times 10^{-2},\ 8 \times 10^{-3}$ from top to bottom.}
\end{figure}

There are rare conditions under which blocks of active sites can merge and exceed the nominal critical charge density of $\bar{n}_c=1.5$: If two of the idealized bubbles studied in Subsection~\ref{subsec:single_active_blocks} converge to $n_A=\bar{n}_c$ on their own, but they end up with only one or two fully-filled sites between them, then the dynamics will allow the two middle sites with $n_i=2$ to become active (see the rightmost gate in Fig.~\ref{fig:markov_circuit}) and the two active blocks will merge into one. This produces one contiguous block of active sites with an average charge density slightly greater than $\bar{n}_c$. We explored this scenario in a series of simulations by evolving two half-filled active blocks of length $L_A \in \{ 8,10,12 \}$ with $n_{i \in A}=1$ separated by $L_A-1$, $L_A$, $L_A+1$, or $L_A+2$ fully-filled sites with $n_{i\in F}=2$ in an otherwise infinite system of $n_i=2$. We ran the dynamics out to $10^{10}$ timesteps and recorded the average charge density and the magnitude of quadrupole moment fluctuations over samples and late times. The results are shown in Fig.~\ref{fig:double_bubble}.

In the cases where the middle block of $n_i=2$ is $L_A$ sites or fewer, the two active blocks are close enough to eventually merge without exceeding the critical density, and the system reaches a featureless equilibrium charge distribution. In contrast, when there are initially $L_A+1$ or $L_A+2$ fully-filled sites in the middle then the active bubbles reach the critical density when there are still one or two frozen sites between them. The dynamics then allows the frozen sites that are left in the middle to become (nominally) active and the two active bubbles coalesce into one that is a total of $1/2$ or $1$ charge over the nominal critical density. In these cases at the latest times the single super-critical active bubble retains a bump of excess charge density at its center---a memory of its initial condition---and thus an active block of this type should not be considered fully thermalizing.  Although the central sites become nominally active, they are not active enough to allow full thermalization. This is also indicated by the significantly suppressed fluctuations of the quadrupole moment of the final active bubble in the super-critical cases (see Fig.~\customref{fig:double_bubble}{(b)}).

From these ``two bubble" simulations we conclude that even though the conditions for creating ``active" blocks with $n_A$ slightly larger than $\bar{n}_c$ will arise in typical systems, these resulting super-critical active blocks do not fully thermalize. Generating these super-critical active blocks requires two growing active blocks to converge to the critical density within one or two sites of each other, and this will happen only rarely as the blocks get large. Sites that thermalize will definitely be active, while the above examples show that active sites do not always thermalize. Another example is the infinitely-repeated charge pattern with spatial period three sites: $...22122122122122...$ where all sites become active, but the average density of $\bar n=5/3$ is well above the critical density and the system does not thermalize. However, we expect that typical states with $\bar n>\bar{n}_c$ will have a nonzero density of frozen sites, and this density thus still can serve as a useful order parameter for the freezing, when averaged over initial states.

\bibliography{main}

\end{document}